\documentclass[12pt,showpacs,showkeys,superscriptaddress,aps]{revtex4}

\usepackage{amssymb}

\usepackage{amsmath,amsthm,amssymb}
\usepackage{epic, eepic}
\newcommand{\maprightu}[1]{%
\smash{\mathop{%
\hbox to 1cm{\rightarrowfill}}\limits^{#1}}}
\newcommand{\maprightd}[1]{%
\smash{\mathop{%
\hbox to 1cm{\rightarrowfill}}\limits_{#1}}}
\newcommand{\mapleftu}[1]{%
\smash{\mathop{%
\hbox to 1cm{\leftarrowfill}}\limits^{#1}}}
\newcommand{\mapleftd}[1]{%
\smash{\mathop{%
\hbox to 1cm{\leftarrowfill}}\limits_{#1}}}
\newcommand{\mapnerssss}[1]{%
\smash{\mathop{%
\hbox to 3cm{\nearrow}}\limits^{#1}}}

\begin{document}

\title{Derivation of Invariant Varieties of Periodic Points from Singularity Confinement in the Case of Toda Map}
\author{Tsukasa YUMIBAYASHI}
\email[email : ]{yumibayashi-tsukasa@ed.tmu.ac.jp}
\affiliation{Department of Physics, Tokyo Metropolitan University,\\
Minamiohsawa 1-1, Hachiohji, Tokyo, 192-0397 Japan}
\author{Satoru SAITO}
\email[email : ]{saito\_ ru@nifty.com}
\affiliation{Hakusan 4-19-10, Midori-ku, Yokohama 226-0006 Japan}
\author{Yuki WAKIMOTO}
\email[email : ]{wakimoto-yuki@ed.tmu.ac.jp}
\affiliation{Department of Physics, Tokyo Metropolitan University,\\
Minamiohsawa 1-1, Hachiohji, Tokyo, 192-0397 Japan}


\begin{abstract}
We have shown in \cite{SS[2]} that the invariant varieties of periodic points (IVPP) of all periods of some higher dimensional rational maps can be derived, iteratively, from the singularity confinement (SC). We generalize this algorithm, in this paper, to apply to any birational map, which has more invariants than the half of the dimension. 
\end{abstract}

\maketitle

\section{Introduction}
Let us consider a map
\begin{equation}
F : x=(x_1,x_2,...,x_d) \rightarrow  X=(X_1,X_2,...,X_d), \qquad x, X \in \hat{\mathbb{C}}^{d}
\label{F}
\end{equation}
where $\hat{\mathbb{C}}:=\mathbb{C} \cup \{ \infty\}\cup \{ 0/0 \}$. Here $\{0/0\}$ denotes a set of indeterminate points, which itself should be clarified in the study.
Since we consider rational maps it is convenient to write them as
\begin{equation}
X_j = \frac{N_j (x)}{D_j (x)}, \qquad (j=1, \dots, d).
\end{equation}
where $N_j(x),\  D_j(x) \in \mathbb{C}[x]$ are irreducible polynomials. We assume that the map is birational. 

Let $j$ be one of $\{1,2,...,d\}$ and denote by $\Sigma_j$ the variety of zero set of $D_j(x)$. We also denote $\Sigma^+:=\bigcup_j \Sigma_j$. The points on $\Sigma^+$ are mapped to $\Lambda(\infty):=F(\Sigma^+)$, which are divergent. But, unless a point of $\Lambda(\infty)$ is a fixed point of the map, there is a possibility that it returns to a finite point after some steps of the map. This is the phenomenon known as singularity confinement (SC) \cite{SC, SC2}. If the points return to a finite region after $m_{sc}$ iteration of the map, we call this number $m_{sc}$, the `steps of SC'. This means that none of $D_j^{(m_{sc}+1)}(\Sigma_j), j=1,2,...,d$ in $F^{(m_{sc}+1)}(\Sigma_j)$ is identically zero, while $F^{(m_{sc})}(\Sigma_j)$ is divergent. It is not difficult to see how this phenomenon takes place \cite{SS[2]}. If $\Sigma^-$ is the zero set of the denominators of the inverse map $F^{-1}$, it is mapped to $F^{-1}(\Sigma^-)\in \Lambda(\infty)$ by the inverse map. Conversely the points on $\Lambda(\infty)$ are mapped back to $\Sigma^-$ by the forward map $F$. From this it is clear that when $F^{(m_{sc})}(\Sigma_j)\in F^{-1}(\Sigma^-)$, it is mapped to $F^{(m_{sc}+1)}(\Sigma_j)\in\Sigma^-$, which is finite. 
\begin{equation}
\begin{array}{ccccccccccccc}
\Sigma_j&\to& F(\Sigma_j)&\to& F^{(2)}(\Sigma_j)&\to& \cdots
&\to& F^{(m_{sc})}(\Sigma_j)&\to& F^{(m_{sc}+1)}(\Sigma_j)&\to&\cdots\\
\cap&&\cap&&\cap&&&&\cap&&\cap&&\\
\Sigma^+&\to&\Lambda(\infty)&\to&\Lambda(\infty)&\to
&\cdots&\to&F^{(-1)}(\Sigma^-)&\to 
&\Sigma^-&\to&\cdots\\
\end{array}
\end{equation}
This is the mechanism that the SC phenomenon undergoes. We should mention here that, although we assumed that $m_{sc}$ is common to all $\Sigma_j$'s, this may not be true in general. Nevertheless all examples we discuss in this paper will satisfy this condition.
\bigskip

Now we assume that the map has $p$ invariants $\{H_1(x), H_2(x),...,H_p(x)\}$. Since the map is constrained on the level set 
\[
V(h)=\Big\{x\Big|H_1(x)=h_1,H_2(x)=h_2,...,H_p(x)=h_p\Big\}
\]
with $h_i$'s being some constants, not all the periodicity conditions $F^{(n)}(x)=x$ of period $n$, but only $d-p$ of them are independent, which we write as
\begin{equation}
\Gamma_j^{(n)}(\xi,h)=0,\qquad j=1,2,...,d-p,\qquad n\ge 2.
\label{Gamma=0}
\end{equation}
Here $\xi=\{\xi_1,\xi_2,...,\xi_{d-p}\}$ is the coordinate which parameterizes the level set $V(h)$. In general (\ref{Gamma=0}) will fix $\xi$ for all $h=\{h_1,h_2,...,h_p\}$, so that points of period $n$ form a set of discrete points. We notice that the fixed points of the map are not counted as periodic points in (\ref{Gamma=0}) and hereafter.

But it might happen that some of the conditions (\ref{Gamma=0}) 
determine relations among the invariants instead of fixing all $\xi$. If $p\ge d/2$, in particular, it is possible that all conditions (\ref{Gamma=0}) of some period $k$ determine relations among the invariants $h$, and do not fix $\xi$ at all. In this case the conditions (\ref{Gamma=0}), which we write as 
\begin{equation}
\gamma_j^{(k)}(h)=0,\qquad j=1,2,...,d-p
\label{gamma}
\end{equation}
specify the geometry of the level set $V(h)$. We notice that all points on the variety defined by
\begin{equation}
v^{(k)}=\Big\{ x\Big|\gamma_j^{(k)}(H(x))=0,\ j=1,2,...,d-p\Big\}
\label{v}
\end{equation}
are points of period $k$.  We call $v^{(k)}$ an invariant variety of periodic points, or IVPP, of period $k$. We have studied in \cite{SS[1]} many rational integrable maps and found that all their periodic points form IVPPs, as far as $p\ge d/2$ is satisfied.

We can prove the following theorem\cite{SS[2], SS[1]}.
\bigskip

\noindent{\bf IVPP theorem:}\quad
{\it When $p\ge d/2$, an IVPP and a set of discrete points of any period can not exist simultaneously in one map.}
\bigskip

Let us present here an outline of the proof , but leaving the details to ref\cite{SS[2]}. Suppose the points of period $k$ are on the IVPP $v^{(k)}$ of (\ref{v}) and the points of period $n(\ne k)$ satisfy (\ref{Gamma=0}) and form a discrete set of points. Since $h$ is free in $\Gamma_j^{(n)}(\xi,h)$ we can always choose the level set to satisfy $\{\gamma_j^{(k)}(h)=0\}$. It means that the discrete set of points of period $n$ are on the variety $v^{(k)}$, which are totally occupied by points of period  $k$. This contradicts to our assumption $n\ne k$.\\

Another important observation in \cite{SS[2]} is the fact that IVPPs of all periods can be derived iteratively once the map is recovered from the SC. It was shown explicitly by studying the 3 dimensional Lotka-Volterra map (3dLV), 
\begin{equation}
F(x)=\left(x_1{1-x_2+x_2x_3\over 1-x_3+x_3x_1},\ 
x_2{1-x_3+x_3x_1\over 1-x_1+x_1x_2},\ 
x_3{1-x_1+x_1x_2\over 1-x_2+x_2x_3}\right),
\label{3dLV}
\end{equation}
which has two invariants
\[
f=x_1x_2x_3-(1-x_1)(1-x_2)(1-x_3),\qquad g=1+(1-x_1)(1-x_2)(1-x_3).
\]
A point 
$
p^{(0)}=\left(x_1,x_2,{1\over 1-x_1}\right)\in \Sigma_1\subset \Sigma^+,
$
which satisfies $
D_1(p^{(0)})=0
$ is mapped iteratively according to
\begin{equation}
p^{(0)}\rightarrow (\infty,0,1)\rightarrow (1,0,\infty)\rightarrow
\left({1\over 1-x_1},x_2,x_1\right)\in \Sigma^-,
\end{equation}
hence the steps of SC is $m_{sc}=3$. 

We notice that, in terms of the invariants, we can write $p^{(0)}$ as
\[
p^{(0)}=\left({f\over f+g}, {g(f+g-1)\over f}, {f+g\over g}\right).
\]
Therefore it is apparent that all $p^{(n)}$'s must be written only by the invariants. In fact we find
\[
p^{(3)}=\left({f+g\over g},{g(f+g-1)\over f},{f\over f+g}\right).
\]
From this expression we see that the denominator $D_1^{(3)}$ of $X_1^{(3)}(p^{(0)})$ is given by $g$. Because $X^{(1)}(p^{(0)})$ is divergent, $g$ must vanish at the points of period 2. In other words the set of points satisfying $g(x)=0$ form an IVPP of period 2. 
We can continue this procedure to obtain all IVPPs of the 3dLV, 
\[
\gamma^{(2)}=g,\quad \gamma^{(3)}=f^2+fg+g^2,\quad
\gamma^{(4)}=f^3+(1-g)(f+g)^3, \quad \cdots
\]
which agree exactly with those derived from the periodicity conditions (\ref{Gamma=0}) directly \cite{SS[1]}.\\

We can apply this algorithm to other maps when the number $p$ of the invariants is $d-1$. In fact we can derive IVPPs of 3d KdV map and 4dLV map in this method. When $p$ is less than $d-1$, a single polynomial of the invariants is not sufficient to determine IVPP of each period. Since we have not studied such cases so far, we must develop a new method. For this purpose we propose in \S 2 a new algorithm and show in \S 3 how it works when $p=d-2$, including the 3 point Toda map.   

\section{Algorithm for general cases}

The key point of the method used in the derivation of IVPPs of the 3dLV map was to parameterize the zero set of the denominators of the map in terms of the invariants. We now ask if there exists a way to derive IVPPs by the SC when $d/2\le p\le d-2$. To answer this question we notice that, in addition to $p^{(0)}\in\Sigma^+$, we need $d-p-1$ other conditions to write down $p^{(0)}$ by the invariants. We would like to propose, in this section, a 
set of such additional conditions and provide an algorithm which enables us to generate all IVPPs from the SC in general cases.

To this end let $F$ be the rational map of (\ref{F}) with $p$ invariants. We assume that $p\ge d/2$ and $m_{sc}$ is finite. Without loss of generality we assume $p^{(0)}\in \Sigma_1$, or equivalently $D_1(p^{(0)})=0$. Now we propose the following algorithm:
\bigskip

\noindent
{\bf Algorithm}

\begin{enumerate}
\item Determination of the initial point $p^{(0)}$:

We impose additional $d-p-1$ conditions $D_1^{(k)}(p^{(0)})=0,\ k=2,3,...,d-p$ and solve 
\begin{eqnarray}
&&\Big\{D_1^{(1)}(x)=0, D_1^{(2)}(x)=0, \cdots, D_1^{(d-p)}(x)=0\nonumber\\
&&\qquad\qquad ,H_1(x)=h_1, H_2(x)=h_2,\cdots, H_p(x)=h_p\Big\} 
\label{conditions}
\end{eqnarray}
for $x$ to determine $p^{(0)}$ by the invariants $h=(h_1,h_2,...,h_p)$.

\item
Generation of IVPPs:

Compute $F^{(k)}(p^{(0)})$ and get
\[
 \big\{D_1^{(k+1)}(p^{(0)}),\ D_1^{(k+2)}(p^{(0)}),\ ...,\ D_1^{(k+d-p)}(p^{(0)})\big\},\qquad k\ge m_{sc}-1
\] 
from which we find $d-p$ irreducible polynomials of $h$, one from each element. They are nothing but $\{\gamma_j^{(k)}(h)\}$ of (\ref{gamma}).
\end{enumerate}
\bigskip

In order to justify our Algorithm we first notice that $D_1^{(2)}(p^{(0)})$ is not identically zero when $D^{(1)}_1(p^{(0)})=0$, because, otherwise, $D_1^{(n)}(p^{(0)})=0$ for all $n\ge 2$, and contradicts to our assumption that $m_{sc}$ is finite. Therefore the conditions (\ref{conditions}) can determine $p^{(0)}$. The second part of the algorithm is apparent because the $k$ period conditions of the map require $D_1^{(j+k)}(p^{(0)})=0$ if $D_1^{(j)}(p^{(0)})=0$ for all $j$.

\section{Application to some maps}

We have already derived the IVPP of period 3 of the 3 point Toda map in \cite{SS[1]}. We would like to mention, however, that  the analysis using computer algebra becomes much harder as the degrees of freedom of the map increases, if we derive the IVPP's directly from the periodicity conditions of the map. We were not able to find IVPP's of periods higher than 3 by using our personal computer.

We apply, in this section, our algorithm in the cases of 4dLV, 5dLV and 3 point Toda map. The 3 point Toda map is related to the 6dLV by a proper transformation of the variables. We study these maps because all invariants have been given explicitly for the LV maps of all dimensions \cite{HT}, so that we can test our algorithm. The number of the invariants of the $d$ dimensional LV map is $p=(d+2)/2$ if $d$ is even while $p=(d+1)/2$ if $d$ is odd \cite{SS[1]}, hence the condition $p\ge d/2$ is satisfied.
  
\subsection{4 dimensional Lotka-Volterra map}

The 4dLV map is given by 
\begin{eqnarray*}
F(x)&=& \left( x_1\frac{1-x_2-x_3+x_2x_3+x_3x_4}{1-x_3-x_4+x_3x_4+x_4x_1} , x_2\frac{1-x_3-x_4+x_3x_4+x_4x_1}{1-x_4-x_1+x_4x_1+x_1x_2} \right. \nonumber \\
&& \left. \quad , x_3\frac{1-x_4-x_1+x_4x_1+x_1x_2}{1-x_1-x_2+x_1x_2+x_2x_3} , x_4\frac{1-x_1-x_2+x_1x_2+x_2x_3}{1-x_2-x_3+x_2x_3+x_3x_4} \right).
\end{eqnarray*}
There are three invariants
\[
\left\{
\begin{array}{ccl}
r&=&x_1x_2x_3x_4 \\
f&=&x_1x_2x_3x_4-(1-x_1)(1-x_2)(1-x_3)(1-x_4) \\
g&=&2-q_1-q_2-q_3-q_4
\end{array}
\right.
\]
where we use the notation $q_j:=x_j(1-x_{j-1}),\ j=1,2,3,4$, and $x_{j+4}=x_j$.

Since $d-p=1$, the single condition $D_1(x)=0$ is sufficient to determine the initial point $p^{(0)}$ on $\Sigma_1^+$. We find
\begin{eqnarray*}
p^{(0)}&=&\left(\frac{(-g^2+g-gf+2f+gN_4)f}{2(gf-g^2r+f^2)}
, -\frac{gf-2gf-f+f^2+fN_4}{2f} , \right. \nonumber \\
&& \quad \left. -\frac{(-gf+2gr+f-f^2+fN_4)g}{2(-gf-g^2r+f^2+g^2f)}
,\frac{-g-2f+g^2+gf+gN_4}{2(g-1)g} \right)
\end{eqnarray*}
where
\[
N_4:=\sqrt{1-2g-2f+g^2+2gf+f^2+4r-4gr}.
\]
$p^{(0)}$ is then mapped to
\begin{eqnarray*}
p^{(0)} &\rightarrow&
\left( \infty , 0 , \frac{f+g-1-N_4}{2(g-1)} , 1\right)\qquad\qquad\qquad\qquad\\
&&\qquad\qquad \rightarrow  
\left( 1 ,\frac{f+g-1+N_4}{2(g-1)} , 0 , \infty \right)
\rightarrow p^{(3)}\rightarrow p^{(4)}\to\cdots,
\end{eqnarray*}
where
\begin{eqnarray*}
p^{(3)}&:=& \left(\frac{-g-2f+g^2+gf+gN_4}{2(g-1)g}
, -\frac{(-gf+2gr+f-f^2+fN_4)g}{2(-gf-g^2r+f^2+g^2f)} , \right. \nonumber \\
&& \quad  \left. -\frac{gf-2gr-f+f^2+fN_4}{2f}
,\frac{(-g^2+g-gf+2f+gN_4)f}{2(gf-g^2r+f^2)} \right),
\label{4dLV SC map}
\end{eqnarray*}
hence the SC step number is 3. If we continue the map further the IVPPs of this map are generated from the denominators $D_1^{(n)}$ of $X_1^{(n)}$, $n=3,4,5,...$ as
\begin{eqnarray}
\gamma^{(2)} &=& g \nonumber\\
\gamma^{(3)} &=& f^2-g^2r+g^2f \nonumber\\
\gamma^{(4)} &=& -2g^2r+g^2f+2f^2\nonumber\\
\gamma^{(5)} &=& r^2fg^6+3f^5g^2+3g^4r^2f^2-3rf^4g^2-4g^4rf^3+f^6-r^3g^6+g^4f^4 \nonumber\\
{\rm etc.}&&
\label{IVPP of 4dLV}
\end{eqnarray}

We notice that a square root singularity appears in the map. 
It should be mentioned that the Gr\"obner basis becomes quite useful to derive the IVPPs in such cases. 

\subsection{5 dimensional Lotka-Volterra map}

We now study the 5dLV map which is the case $p=d-2$, so that
we must apply our new Algorithm in \S 2. 

The 5dLV map is given by
\[
F(x)=\left(x_1{D_5\over D_1},\ x_2{D_1\over D_2},\ x_3{D_2\over D_3},\ x_4{D_3\over D_4},\ x_5{D_4\over D_5}\right)
\]
where we used the notation
\[
D_j:=(1-x_{j+2})\Big(x_jx_{j+4}+(1-x_{j+3})(1-x_{j+4})\Big)
+x_{j+2}x_{j+3}x_{j+4}x_j,\quad j=1,2,...,5,
\]
with $x_{j+5}=x_j$. This map has three invariants,
\begin{equation}
\left\{
\begin{array}{ccl}
r &=& x_1x_2x_3x_4x_5\\
f &=&x_1x_2x_3x_4x_5- (1-x_1)(1-x_2)(1-x_3)(1-x_4)(1-x_5)\\
g &=&2-q_1-q_2-q_3-q_4-q_5+x_1x_2x_3x_4x_5.\\
\end{array}\right.
\end{equation}
Since $d-p=2$ in this map we impose, according to the first step of the Algorithm, two conditions $D_1(x)=0$ and $D_1^{(2)}(x)=0$ to obtain the initial point $p^{(0)}$ on $\Sigma_1$,
\[
p^{(0)}=\left({f^2g\over B},r{B-fA\over -f^3},{B-fg^2\over B-fA}
,{(B-gf^2)f\over (B-fg^2)(f-g)}, {B(g-f)\over (B-gf^2)g}\right)
\]
where
\[
A:=f^2-gf+g^2, \quad B:=rA-f^2.
\]
The second step of the Algorithm shows that the SC map undergoes as
\[
p^{(0)}\rightarrow \left(\infty,0,{g\over g-f},{f\over g},1\right)
\rightarrow \left({0\over 0},0,1,{0\over 0},{0\over 0}\right)\rightarrow \left(1,0,{0\over 0},{0\over 0},{0\over 0}\right)\qquad
\]
\begin{equation}
\qquad\qquad
\rightarrow\left({g\over g-f},0,\infty, 1,{f\over g}\right)
\rightarrow p^{(5)}\rightarrow p^{(6)}\rightarrow\cdots
\label{5dLV SC}
\end{equation}
where
\[
p^{(5)}=\left({B-fg^2\over B-fA},r{B-fA\over -f^3},
{f^2g\over B},{B(g-f)\over (B-gf^2)g},
{(B-gf^2)f\over (B-fg^2)(f-g)}\right), \quad etc..
\]
Hence the steps of SC is 5.  In (\ref{5dLV SC}), $0/0$ means that the denominator and the numerator of the component become zero separately, so that we can not determine its value. In other words the point is indeterminate. 

We can derive IVPPs of periods $k\ge m_{sc}-1=4$, following to the second step of the Algorithm. Namely we should be able to find two irreducible polynomial functions 
$\gamma_1^{(k)}$ and $\gamma_2^{(k)}$ of the invariants from each pair  $\{D_1^{(k+1)}, D_1^{(k+2)}\}$, $k=4,5,6,...$. Then their intersection $v^{(k)}=\{ \gamma_1^{(k)}=0,\ \gamma_2^{(k)}=0\}$ forms the IVPP of period $k$ of this map.  The results are listed in (\ref{gammaof period5}).  We should mention that the final expression of $\gamma^{(k)}_{1,2}$'s are given by means of the Gr\"obner basis in the list.

Since our Algorithm does not tell how to get IVPPs of period  $k<m_{sc}-1$, we must find IVPP of period 3 separately. It is done easily if we notice that $D_1^{(4)}$ and $D_1^{(5)}$ are not identically zero while $D_1^{(1)}$ and $D_1^{(2)}$ have been chosen to vanish. The result of this case is also included in the list below.
\begin{eqnarray}
\gamma_1^{(3)} &=& g-f ,\quad \gamma_2^{(3)} = r-f-1 \nonumber\\
\gamma_1^{(4)}&=&2f^2-2fg+g^2,\quad \gamma_2^{(4)}=f^2+f-g-rf\nonumber\\
\gamma_1^{(5)}&=&f^4-3gf^3+4f^2g^2-2fg^3+g^4\nonumber\\
\gamma_2^{(5)}&=&-2f^3+2f^2-rf^2+(6f^2-f)g-(3f-1)g^2+2g^3\nonumber\\
{\rm etc.}&&
\label{gammaof period5}
\end{eqnarray}

\subsection{3 point Toda map}

The 6dLV map, which is much more complicated than the 5dLV map, becomes simpler if we transform it to the 3 point Toda map by the Miura transformation \cite{HT}:
\[
x=(1-x_1)(1-x_2),\quad y=(1-x_3)(1-x_4),\quad z=(1-x_5)(1-x_6),
\]
\[
u=x_2x_3,\quad v=x_4x_5,\quad w=x_6x_1.
\]
After the transformation the 3 point Toda map is given by
\begin{eqnarray}
F(x,y,z,u,v,w)
&=& \left( y\frac{zu+zx+wu}{yw+yz+vw} ,z\frac{xv+xy+uv}{zu+zx+wu}
 , \right. \nonumber \\
&& \quad \left. x\frac{yw+yz+vw}{xv+xy+uv}, u\frac{yw+yz+vw}{zu+zx+wu} ,  \right. \nonumber \\
&& \quad \quad \left. v\frac{zu+zx+wu}{xv+xy+uv} ,w\frac{xv+xy+uv}{yw+yz+vw} \right).
\end{eqnarray}
This map has four invariants,
\begin{equation}
\left\{
\begin{array}{ccl}
r&=&xyz\\
t&=&x+y+z+u+v+w\\
f&=&xy+yz+zx+uv+vw+wu+xv+yw+zu \\
g&=&uvw-xyz.\\
\end{array}
\right.
\end{equation}

According to the first step of the Algorithm, we parametrize $\Sigma_x$ by using  $D_x^{(2)}$,  and obtain
\begin{eqnarray}
p^{(0)}&=&\left(\frac{r(-g^2t+gf^2+f^2r)}{g^3}
, \frac{g^2f}{-g^2t+f^2r} , \right. \nonumber \\
&& \quad \left. \frac{(-g^2t+f^2r)g}{(-g^2t+gf^2+f^2r)f}, 
  -\frac{(r+g)(-g^2t+f^2r)}{g^3}, \right. \nonumber \\
&& \quad \quad \left. -\frac{g(-g^2t+gf^2+f^2r)}{(-g^2t+f^2r)f}
, \frac{g^2f}{-g^2t+gf^2+f^2r}
 \right) .
\label{Toda p^0} 
\end{eqnarray}
We can proceed easily the second step of the Algorithm to see how the SC map undergoes
\begin{eqnarray}
p^{(0)} &\rightarrow&
\left( \infty , -\frac{g}{f} , 0 , 0 , \frac{g}{f} , \infty \right) \rightarrow
\left( {0\over 0} , 0 , {0\over 0}, 0 , {0\over 0}, {0\over 0} \right)  \nonumber \\
&\rightarrow& \left( 0, {0\over 0}, {0\over 0}, 0 ,{0\over 0}, {0\over 0} \right) \rightarrow
\left( -\frac{g}{f} , \infty , 0 ,0 , \infty , \frac{g}{f} \right) \nonumber \\
&\rightarrow& p^{(5)}\rightarrow p^{(6)}\to \cdots \label{Toda SC map}
\end{eqnarray}
where
\begin{eqnarray}
p^{(5)}&=& \left( \frac{g^2f}{-g^2t+f^2r}
, \frac{r(-g^2t+gf^2+f^2r)}{g^3}, \right. \nonumber \\
&& \quad \left. \frac{(-g^2t+f^2r)g}{(-g^2t+gf^2+f^2r)f}, -\frac{(r+g)(-g^2t+f^2r)}{g^3}, \right. \nonumber \\
&& \quad \quad \left. \frac{g^2f}{-g^2t+gf^2+f^2r}
, -\frac{g(-g^2t+gf^2+f^2r)}{(-g^2t+f^2r)f}
 \right),\quad {\rm etc.}. 
\label{p^5}
\end{eqnarray}
Hence $m_{sc}=5$ in this map.

To find the IVPPs we notice that $\{D_x^{(k)}, D_x^{(k+1)}\},\ k=3,4,5,...$ are not identically zero while $ D_x^{(1)}=D_x^{(2)}=0$ have been imposed. We thus obtain the following list of IVPPs.
\begin{eqnarray*}
\gamma_1^{(3)}&=&f,\qquad \gamma_2^{(3)}=tg^2\\
\gamma_1^{(4)}&=&g^2t-f^2r,\qquad \gamma_2^{(4)}=tf+g\\
\gamma_1^{(5)}&=&-2f^3r^2+2frtg^2-f^3rg-g^4,\quad
\gamma_2^{(5)}=4f^3rg^3+f^6r^2-g^6\\
\gamma_1^{(6)}&=&5f^4r^2+5gf^4r-6g^2f^2tr+g^2f^4-3g^3f^2t
+2g^4f+g^4t^2,\\
\gamma_2^{(6)}&=&4f^6+3t^2f^2g^2-4f^3g^2+g^4\\
{\rm etc.}&&
\end{eqnarray*}

Finally we remark that, since we have chosen the function $D_x^{(2)}$ as the additional condition for the parametrization of $\Sigma_x$, the steps of the SC increased from 3 to 5. This change precludes the decision of the IVPPs of period 2. However we can always impose periodicity conditions to find the IVPP of each period separately. For instance, in the period 2 case, we can manipulate the periodicity conditions directly, and find, after battle with computer, that some points on the intersection of the following three hypersurfaces
\begin{eqnarray*}
xyu^3-3xyz(x+y-z+u)u-z(x^2+yz)(y^2+zx)&=&0\\
yzv^3-3xyz(y+z-x+v)v-x(y^2+zx)(z^2+xy)&=&0\\
zxw^3-3xyz(z+x-y+w)w-y(z^2+xy)(x^2+yz)&=&0
\end{eqnarray*}
satisfy the period 2 conditions.

\section{Conclusion}

We have explored in this paper how IVPPs are generated by the SC. This mechanism of generation of infinite sequence of algebraic varieties is quite impressive. In order to understand 
this phenomenon we have investigated in \cite{S} the Hirota-Miwa equation itself, from which the KP hierarchy is derived, by means of the theory of derived category. It is shown that the localization of the map is associated with this phenomenon at least in the 3dLV case. Extension to higher dimensional maps is  under investigation. 





\bibliographystyle{elsarticle-num}
\bibliography{<your-bib-database>}



\end{document}